\documentclass[conference]{IEEEtran}
\IEEEoverridecommandlockouts
\usepackage{cite}
\usepackage{amsmath,amssymb,amsfonts}
\usepackage{graphicx}
\usepackage{textcomp}
\usepackage[dvipsnames]{xcolor}
\usepackage{soul}
\usepackage[ruled]{algorithm2e}
\usepackage{algpseudocode}
\usepackage{tikz}
\usepackage{braket}
\usetikzlibrary{quantikz}
\usepackage{multicol}
\usepackage[caption=false]{subfig}
\usepackage{xspace}

\usepackage{tikz}
\usetikzlibrary{fit,calc}

\makeatletter
\newcommand{\removelatexerrorr}{\let\@latex@error\@gobble}
\makeatother

\colorlet{pink}{red!40}
\colorlet{blue}{cyan!60}
\colorlet{green}{green!60}

\def\BibTeX{{\rm B\kern-.05em{\sc i\kern-.025em b}\kern-.08em
    T\kern-.1667em\lower.7ex\hbox{E}\kern-.125emX}}

\newcommand{\CartPole}{\mbox{CartPole}\xspace}
\newcommand{\CartPoleEnv}{\mbox{\texttt{CartPole-v1}} environment\xspace}

\newcommand{\removelatexerror}{\let\@latex@error\@gobble}

\begin{document}
\bstctlcite{IEEEexample:BSTcontrol}

\title{Photonic Quantum Policy Learning in OpenAI Gym\\
}

\author{
\IEEEauthorblockN{D\'aniel Nagy}
\IEEEauthorblockA{
    \textit{Wigner Research Centre for Physics and}\\
    \textit{Ericsson Research } \\
            Budapest, Hungary \\
    nagy.dani@wigner.hu
}
\and
\IEEEauthorblockN{Zsolt Tabi}
\IEEEauthorblockA{
    \textit{Ericsson Hungary and}\\
    \textit{E\"otv\"os Lor\'and University}\\
    Budapest, Hungary \\
    zsolt.tabi@ericsson.com
}
\and
\IEEEauthorblockN{P\'eter H\'aga}
\IEEEauthorblockA{
    \textit{Ericsson Research} \\
    Budapest, Hungary \\
    \hspace*{2mm} peter.haga@ericsson.com \hspace*{2mm}
}
\and 
\IEEEauthorblockN{\hspace*{45mm}Zs\'ofia Kallus}
\IEEEauthorblockA{
    \hspace*{45mm}\textit{Ericsson Research} \\
    \hspace*{45mm}Budapest, Hungary \\
    \hspace*{45mm} zsofia.kallus@ericsson.com} \\
\and
\IEEEauthorblockN{Zolt\'an Zimbor\'as}
    \IEEEauthorblockA{\textit{Wigner Research Centre for Physics and} \\
    \textit{ MTA-BME Lend\"ulet QIT Research Group}\\
    Budapest, Hungary \\
    zimboras.zoltan@wigner.hu
}
}

\maketitle

\begin{abstract}
In recent years, near-term noisy intermediate scale quantum (NISQ) computing devices have become available. One of the most promising application areas to leverage such NISQ quantum computer prototypes is quantum machine learning.
While quantum neural networks are widely studied for supervised learning, quantum reinforcement learning is still just an emerging field of this area.
To solve a classical continuous control problem, we use a continuous-variable quantum machine learning approach. We introduce proximal policy optimization for photonic variational quantum agents and also study the effect of the data re-uploading. We present performance assessment via empirical study using Strawberry Fields, a photonic simulator Fock backend and a hybrid training framework connected to an OpenAI Gym environment and TensorFlow. For the restricted CartPole problem, the two variations of the photonic policy learning achieve comparable performance levels and a faster convergence than the baseline classical neural network of same number of trainable parameters.
\end{abstract}


\section{Introduction}

Exceptional progress has been made in the last years on the path to the development of fully functional quantum processors, opening a new era in quantum computing. With state-of-the-art quantum computers reaching a scale of 50-100 qubits, the scientific community is approaching a regime in which quantum systems may give rise to computational advantage, as was demonstrated in the sampling experiments by Google \cite{arute2019quantum} and the USTC team \cite{zhong2020quantum}. 
These results pave the way towards a wide availability of noisy intermediate-scale quantum (NISQ)  devices for researchers to work on potential applications \cite{Preskill2018quantumcomputingin}. One of the leading candidates for practical quantum advantage, right next to the simulation of quantum chemistry and many-body systems \cite{arute2020hartree,bauer2016hybrid,cade2020strategies, king2021scaling}, are variational optimization methods \cite{farhi2014quantum,harrigan2021quantum,kerenidis2019quantum} and Quantum Machine Learning (QML) \cite{biamonte2017quantum,ciliberto2018quantum,liu2018quantum, schuld2019quantum,havlivcek2019supervised, huang2021information}. These methods are inherently noise resistant and land to a natural adaptation to parametric quantum circuits. 

Integrated photonics offers one of the experimentally most feasible platforms for scalable quantum computation \cite{qiang2018large, wang2020integrated, arrazola2021quantum, bombin2021interleaving}. 
Besides using this architecture for building logical qubits for standard quantum computation \cite{knill2001scheme},  photonic circuits can also realize the special case of continuous-variable universal quantum computation \cite{lloyd1999quantum,braunstein2005quantum, serafini2017quantum, arrazola2019quantum}. These characteristics motivate the current active research on photonic QML \cite{lau2017quantum,CVQNNLLoyd, bartkiewicz2020experimental, javsek2019experimental}. 
Moreover, it should be mentioned that even today, photonic (classical) computation prototypes are used as AI accelerators due to high-speed and low-energy implementation of tensor calculus \cite{ramey2020silicon,harris2020accelerating}.

The paper is organized as follows. In the next section, Reinforcement Learning is reviewed. In Sec.~\ref{sec:photonic_policy_learning}, we present the photonic quantum policy learning algorithm and the CartPole problem that will be the subject to our empirical tests. The software tools and methods used in our experiments are described in Sec.~\ref{sec:hybrid_training}. The aforementioned empirical results are described in detail in Sec.~\ref{sec:empirical_results}. Finally, Sec.~\ref{sec:conclusion} summarizes our findings and gives an outlook on some potential research directions. 

\section{Reinforcement Learning}
\label{sec:RL}
Reinforcement Learning (RL) is sub-field of Machine Learning in which one would like to train an \textit{agent} to optimally perform some \textit{actions} in an \textit{environment}, getting \textit{rewards} as feedback to improve its behavior. 
The general goal is to improve the \textit{cumulative rewards} throughout the training steps by interacting with the environment. RL algorithms are either model-based, learning a predictive model of the environment, or model-free, i.e., learning a control policy directly. Model-free algorithms are either policy-based (actor), value-based (critic), or actor-critic (AC) methods representing the policy function independently of the value function.

The policy function $\pi$ returns a probability distribution over the possible actions in a given state. The state-dependent on-policy value function, $V^\pi$ calculates expected return starting at a given state and running the policy from there on. The critic hence evaluates the actions, and the actor updates the policy gradient based on the critic's feedback \cite{sutton1999policy}. In this work, we implement the actor and the critic via two separate neural networks and introduce a generalized hybrid model with a classical critic and a quantum actor network.

We focus on state-of-the-art algorithms which excel in learning to solve continuous control problems while they are simple to implement. Policy gradient methods overcome limitations of traditional approaches by optimizing the policies directly. Instead of learning a value function, they define a parameterized policy $\pi_{\theta}$ and optimize $\theta$ by maximizing the expected returns. 

Instead of a combined value and policy network, we separate the policy function into a parametric QNN circuit.
A classical NN is used to approximate the value function, $V^{\pi}_{a}$, where
$a$ represents the trainable parameters of the value function, which are separate from the policy parameters~$\theta$. This value function is being optimized by calculating  the total discounted reward collected by the policy,
\begin{equation}
    V_{a,\textrm{targ}}^{\pi}(s_t) = \sum\limits_{l=0}^{T-t} \gamma^lr_{t+l} \enspace,
\end{equation}
and then optimizing the parameters $a$ via a mean-squared error loss,
$L^{VF} = \underset{t}{\LARGE{\mathbb E}} \left[\left(V^{\pi}_{a}(s_t) - V_{a,\textrm{targ}}^{\pi}(s_t)\right)^2\right]$.
Here, $r_{t}$ denotes the reward collected at timestep $t$, and $\gamma$ is the discount factor, typically close to $1$.

For the policy circuit, we use policy gradient optimization, naturally adaptable to quantum circuits. The objective function's gradient is the policy gradient estimator, and its basic form is defined by $L^{PG}\left(\theta\right) = \mathbb{\hat{E}}_t \left[\log \pi_\theta(a_t|s_t\hat A_t) \right]$, where $\hat{A}_t$ denotes the \textit{estimated advantage} at time step $t$:
\begin{equation}
    \hat A_t = \sum\limits_{l=0}^{T-t-1} (\gamma\lambda)^{l}\delta_{t+l},
\end{equation}
\noindent and $\delta_t = r_t + \gamma V^{\pi}(s_{t+1}) - V^{\pi}(s_t)$, where $\lambda$ is the hyperparameter of Generalized Advantage Estimation (GAE) \cite{Schulmanetal_ICLR2016}.

However, the problem with the basic $L^{PG}$ is that it often leads to destructively large policy updates. This is usually mitigated by the use of trust region and proximal policy optimization methods.
We used a combination of both in the following way.

Our main goal was to adapt the Proximal Policy Optimization (PPO) algorithm introduced in Ref.~\cite{schulman2017proximal} using a clipped surrogate objective function.
A $\text{clip}(\cdot)$ function is used for clipping the probability ratios
$r_t(\theta)=\pi_{\theta}/\pi_{\theta_{\textrm{old}}}$ so that moving it outside of
the proximity region $[1-\epsilon, 1+\epsilon]$ defined by $\epsilon$ hyperparameter does not have an incentive. This gives 
\begin{equation}
L^{\text{CLIP}}(\theta) = \mathbb{\hat{E}}_t \left[ \min\left( r_t(\theta)\hat A_t, \textrm{clip}\left(r_t(\theta),1{-}\epsilon,1{+}\epsilon\right)\hat A_t\right) \right] .
\end{equation}
Furthermore, Trust Region methods \cite{schulman2020trust} mitigated the problem by introducing a KL divergence penalty term $D^{KL}(\pi_{\theta}||\pi_{\theta_{\textrm{old}}})$, so that large policy updates are suppressed. The strength is tuned via a hyperparameter, $\beta$, which we chose to be set 
to a weak coupling.  
We also keep the entropy bonus term $S[\pi_\theta]$ \cite{schulman2017proximal}, coupled by the hyperparameter $c_2$, to ensure sufficient exploration. 
Finally, we include an $L_2$ regularization term on the parameters of the active gates (i.e., displacement and squeezing) with a hyperparameter factor $\alpha$, so that the purity of the quantum state $\textrm{Tr}[\rho^2]$ remains high. See details of the circuit in Sec.~\ref{sec:photonic_policy_learning}.

Putting it all together, keeping the notation of \cite{schulman2017proximal}, the terms of the final objective function $L_t^{\text{PPO}}(\theta)$ to be maximized over $\theta$ for our photonic PPO agent takes the following form:
\begin{equation}
\mathbb{\hat{E}}_t \left[ L_t^{\text{CLIP}}(\theta) 
{-} \beta D^{KL}(\pi_{\theta}||\pi_{\theta_{\textrm{old}}}) {+} c_2 S \left[ \pi_{\theta} \right] (s_t) \right] {-} \alpha L_2.
\label{eq:combined_loss}
\end{equation}

The fine-tuning of the hyperparameters introduced has been performed in a minimal, greedy search, 
and a set of fixed parameters were used from the original paper, detailed in the next section.

\section{Photonic Quantum Policy Learning}
\label{sec:photonic_policy_learning}

There are known examples of QRL methods introduced for classical problems, such as \cite{dunjko2017advances,albarran2018measurement, benedetti2019parameterized,  chen2020variational, lockwood2020reinforcement, hamann2020quantum, skolik2021quantum,jerbi2021quantum}. Among these, continuous-variable QRL models are still scarce \cite{hu2019reinforcement, wu2020quantum, flamini2020photonic,saggio2021experimental, lamata2021quantum}. Here, we show a new example for photonic AC learning leveraging the PPO framework presented in Sec.~\ref{sec:RL}.

\subsection{Photonic Proximal Policy Optimization}
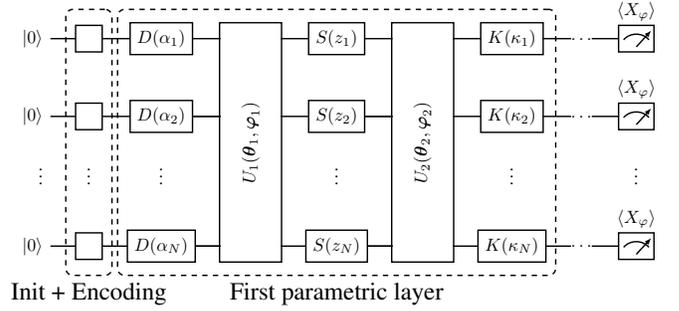
\begin{figure}[t]
\newcommand{\GG}[2]{\gategroup[#1,steps=#2,style={dashed,rounded corners,inner xsep=2pt},background,label style={label position=below,anchor=north,yshift=-0.2cm}}
\newcommand{\Meter}[1]{\meter{$\langle #1 \rangle$}}
\centering
\scalebox{.67}{\begin{quantikz}[transparent]
\lstick{$\ket{0}$} & \gate{\ \ } \GG{4}{1}] {{ \Large Init + Encoding }}
                   & \gate{{D(\alpha_1)}} \GG{4}{5}] {{ \Large First parametric layer }} 
                   & \gate[4,nwires=3,style={text width=35pt},label style={rotate=90}]{\hspace{-0.35cm}U_1(\boldsymbol{\theta}_1, \boldsymbol{\varphi}_1) \hspace{-0.35cm}}
                   & \gate{S(z_1)}
                   & \gate[4,nwires=3,style={text width=35pt},label style={rotate=90}]{\hspace{-0.35cm}U_2(\boldsymbol{\theta}_2, \boldsymbol{\varphi}_2) \hspace{-0.35cm}}
                   & \gate{ K(\kappa_1)}
                   & \qw \ldots
                   & \Meter{ X_\varphi} \\
\lstick{$\ket{0}$} & \gate{\ \ }
                   & \gate{D(\alpha_2)}
                   & \qw 
                   & \gate{S(z_2)}
                   & \qw 
                   & \gate{K(\kappa_2)}
                   & \qw \ldots
                   & \Meter{X_\varphi} \\
\lstick{\vdots}    & \vdots
                   & \vdots
                   & 
                   & \vdots
                   & 
                   & \vdots
                   & 
                   & \vdots
                   & \\
\lstick{$\ket{0}$} & \gate{\ \ }
                   & \gate{D(\alpha_N)}
                   & \qw 
                   & \gate{S(z_N)}
                   & \qw 
                   & \gate{K(\kappa_N)}
                   & \qw \ldots
                   & \Meter{X_\varphi}
\end{quantikz}}
\vspace{0.1cm}
\caption{The QNN architecture. The circuit is composed of a combined initialization and encoding layer and a number of paramateric layers.  The parametric layers consist of displacement gates, interferometers, squeezing and  Kerr gates.}
\label{fig:qnn_gen}
\end{figure}

\begin{figure*}[t]
\newcommand{\GG}[2]{\gategroup[#1,steps=#2,style={dashed,rounded corners,inner xsep=2pt},background,label style={label position=below,anchor=north,yshift=-0.2cm}]}
\newcommand{\Meter}[1]{\meter{$\langle #1 \rangle$}}
\centering
\subfloat[Ansatz circuit with single displacement data encoding]{\scalebox{.8}{
    \begin{quantikz}[transparent]
    \lstick{$\ket{0}$} & \gate{S(0.5)} \GG{2}{1} {{ Init }} 
                   & \gate{D(x_1)} \GG{2}{1} {{ Encoding }}
                   & \gate[2]{BS} \GG{2}{7} {{ First layer }} 
                   & \gate{D}
                   & \gate{R}
                   & \gate[2]{BS} 
                   & \gate{S}
                   & \gate{R}
                   & \gate{K}
                   & \qw \ldots
                   & \Meter{X_1} \\
    \lstick{$\ket{0}$} & \gate{S(0.5)}
                   & \gate{D(x_2)}
                   & \qw 
                   & \gate{D}
                   & \gate{R}
                   & \qw 
                   & \gate{S}
                   & \gate{R}
                   & \gate{K}
                   & \qw \ldots
                   & \Meter{X_2}
    \end{quantikz}}
\label{subfig:qnn321a}}
\vspace{0.1cm}

\subfloat[Ansatz circuit with repeated displacement data encoding (re-uploading) in each layer]{\scalebox{.8}{
    \begin{quantikz}[transparent]
    \lstick{$\ket{0}$} & \gate{S(0.5)} \GG{2}{1} {{ Init }} 
               & \gate{D(x_1)} \GG{2}{8} {{ First layer including encoding }} 
               & \gate[2]{BS}
               & \gate{D}
               & \gate{R}
               & \gate[2]{BS} 
               & \gate{S}
               & \gate{R}
               & \gate{K}
               & \qw \ldots
               & \Meter{X_1} \\
    \lstick{$\ket{0}$} & \gate{S(0.5)}
               & \gate{D(x_2)}
               & \qw 
               & \gate{D}
               & \gate{R}
               & \qw 
               & \gate{S}
               & \gate{R}
               & \gate{K}
               & \qw \ldots
               & \Meter{X_2}
    \end{quantikz}}
\label{subfig:qnn321b}}
\vspace{0.1cm}
\caption{Policy circuit Ans\"atze. (a) presents the Ansatz circuit in which the initialization of parameters and their displacement data encoding is applied only at the beginning of the circuit followed by the first layer of the quantum neural network. (b) presents the initialization and the first layer of the modified circuit allowing the data re-uploading in each layer. In this version the displacement data encoding is applied before every quantum layer. By this re-application of the data encoding operators, 
one can achieve significant boosts in the performance of the quantum neural network.
Note: we applied three QNN layers during the performance evaluation of both cases.
}
\label{fig:qnn32}
\end{figure*}
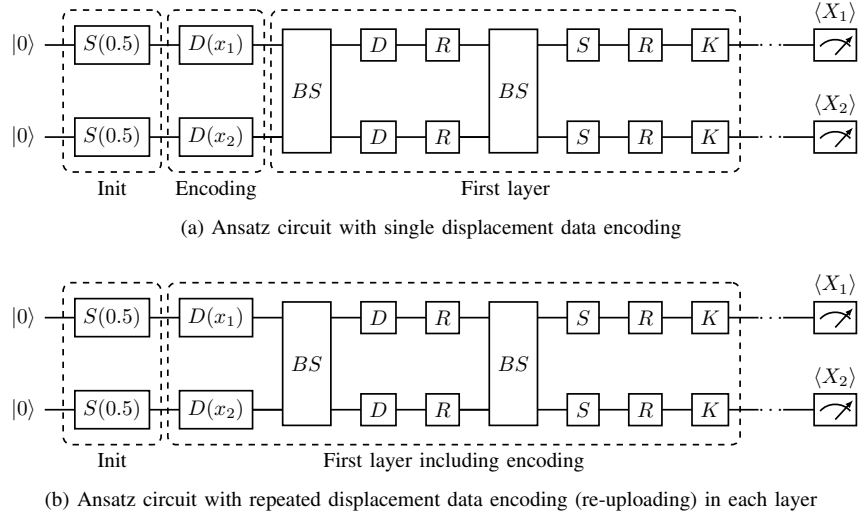

Among the policy optimization methods PPO is one of the best performing algorithms for continuous control problems. Here we introduce for the first time a quantum generalization of PPO leveraging continuous variable quantum neural networks (QNNs) \cite{CVQNNLLoyd}, which we name {\it photonic PPO}. A standard photonic QNN architecture, depicted on Fig.~\ref{fig:qnn_gen},  is composed of an initialization and encoding layer followed by a number of parametric layers each composed of single-qumode displacement gates, an active linear optical circuit element (two interferometers and squeezing gates in between) and single-qumode nonlinear Kerr gates.  Note that the previous set of gates is proven to be universal \cite{lloyd1999quantum,oszmaniec2017universal}, i.e., the effect of any unitary $\hat U$ can be approximated by consecutively applying these gates. Therefore, our Ansatz can approximate (to a certain level due to the limited number of layers) any unitary mapping between the initial state, and the output state. After the parametric layers, a measurement is performed, which is usually taken to be the measurement of the single-mode 
quadrature operators $X_\varphi = X \cos (\varphi) + P \sin (\varphi)$. 

To implement the PPO quantum algorithm with QNNs, one can encode the continuous environment features as displacement, rotation or squeezing parameters in each qumode. 
The parametric layers are initialized with random parameter values, which are updated by the PPO algorithm in each step.
The measured quadrature expectation values at time $t$ determine the updated policy probability values for the environment $s_t$, $\pi_{\theta}( \cdot | s_t )$. In other words, the policy $\pi_{\theta}(\cdot|s_t)$ is defined by the QNN unitary  $\hat U(\theta;s_t)$, which depends on the encoded environment features $s_t$ and 
the learnable parameters $\theta$. The updates of the parameters $\theta$ are done through the optimization of a  PPO loss function of the form presented in Eq.~\eqref{eq:combined_loss}.

Next, we discuss a concrete control task, the \CartPole{} problem and then its solution by a quantum photonic PPO.

\subsection{Photonic PPO for the \CartPole{} problem}
\label{subsec:pppocart}

The well-known \CartPole{} control problem \cite{selfridge1985training} encapsulates the abstraction of balancing a \textit{pole} on a cart which moves along a line. 
The cart can be moved in either direction by one unit. 
The pole starts in $90^\circ$ angle and the goal is to balance the pole so that it does not fall over more than $15^{\circ}$. 
For every time step, if the goal is met, one reward unit is achieved by the agent.

Let us now describe our proposal  for a photonic PPO applied to the \CartPole{} problem. 
Since the problem features are continuous, it is straightforward to use the photonic QNN architecture, as it is also operating with continuous variables. The original problem has four environmental parameters: the pole angle, the angular velocity, the cart position, and the cart velocity. As a simplification, we restrict our investigations to two environmental parameters, the pole angle, and its angular velocity. We encode these two features into two qumodes in the following way: 
Initially, 
the squeezing operation $\hat S(r=0.5, \phi=0)$  is applied on each qumode to the vacuum to generate the following state:
\begin{equation}
   \ket{\psi_0} = \hat S_1\left(\tfrac{1}{2},0\right)\otimes \hat S_2\left(\tfrac{1}{2},0\right) \ket{00}.
\end{equation}
Then we proceed to encode the restricted state of the environment into the quantum state. The pole angle $\varphi$ and angular velocity $\omega$ is mapped into the qumodes with displacement gates as
\begin{equation}
    \ket\psi = \left(\hat D_1\left(x_1,\frac{\pi}{2}\right) \otimes \hat D_2\left(x_2,\frac{\pi}{2}\right)\right)\ket{\psi_0},
\end{equation}
where
\begin{equation}
    x_k = \begin{cases}
    + \frac{4}{\pi} |\arctan(s_k)|^{1/3},~s_k > 0\\
    - \frac{4}{\pi} |\arctan(s_k)|^{1/3},~s_k < 0
    \end{cases},~ k\in\{1,2\}.
    \label{eq:feature_transform}
\end{equation}
In the above formula, $s_1=\varphi$ is the angle of the pole and $s_2=\omega$ is its angular velocity. In other words, these displacement operators encode the classical data into the quantum state by shifting the 
previously squeezed state along the $P_k$-axis by the value of $x_k$.
We chose to apply the transformation \eqref{eq:feature_transform}  on the environment descriptors, 
since $x_k \in [-2, 2]$ is necessary in order to remain in a
suitable displacement range.

After the data encoding, a number of quantum neural network layers are applied with tunable parameters as depicted in Fig.~\ref{fig:qnn32}.
At the end of the circuit a homodyne measurement is performed on the state  $\ket{\psi_{\textrm{out}}}$ 
to obtain the expectation values of the quadrature operators $P_1$ and $P_2$.
Using the measurement results (after enough number of circuit runs or shots) one can estimate the expectation values $\langle P_1\rangle$, $\langle P_2 \rangle$.
We then use these expectation values to sample actions according to the $\tau$-modified softmax probability distribution
\begin{align}
\begin{split}
    \textrm{Pr}(a=0) &= \frac{e^{ \langle P_1\rangle/\tau}}{e^{\langle P_1\rangle/\tau} + e^{ \langle P_2\rangle/\tau}},\\[0.1cm]
    \textrm{Pr}(a=1) &= \frac{e^{ \langle P_2\rangle/\tau}}{e^{\langle P_1\rangle/\tau} + e^{\langle P_2\rangle/\tau}}.
    \end{split}
\end{align}

Finally, it is important to note that the displacement data encoding can be applied before every quantum layer, not only at the beginning of the circuit. This re-application of the data encoding operators is a technique called data re-uploading, and it usually boosts the performance of quantum neural networks.
Fig. \ref{subfig:qnn321a} shows our Ansatz circuit without data re-uploading, while Fig. \ref{subfig:qnn321b} shows the same QNN architecture, but data re-uploading is applied.

\section{Hybrid Training for Photonic PPO}
\label{sec:hybrid_training}

\begin{figure*}[tb]
\centering
\includegraphics[trim={0 3cm 0 0.75},clip,width=0.7\textwidth]{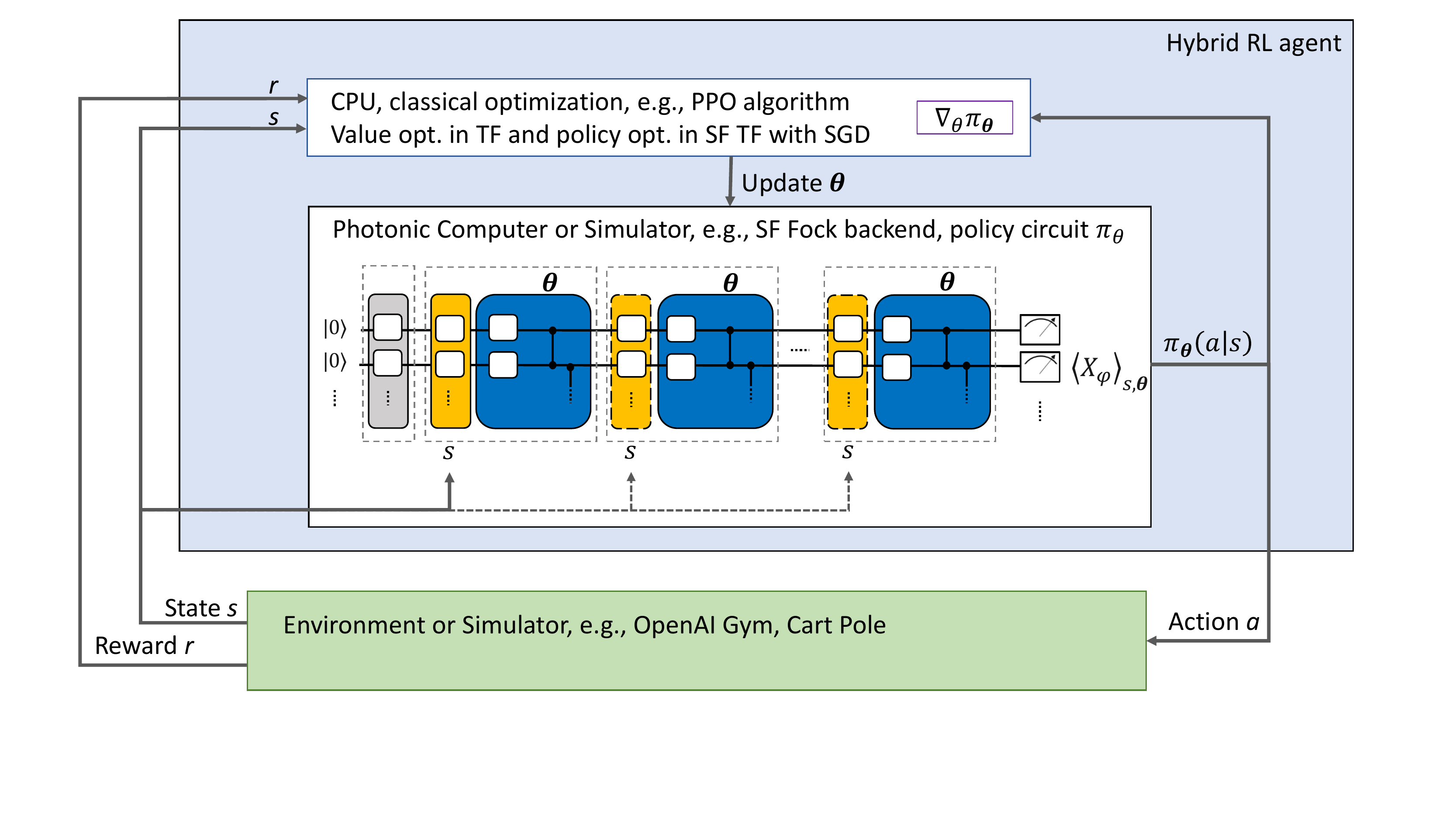}
\caption{Illustration of the high-level hybrid training architecture. The agent sends actions to be executed in an external environment or simulator. In response, a reward $r$ and environment state observation $s$ are reported. The agent's photonic quantum circuit is composed of an initialization layer followed by either a single state encoding and repeated control policy layers, or with data re-uploading inserted before each policy layer. The parametric circuit is then trained via classical optimization using homodyne measurements as the quantum output layer.}
\label{fig:arch}
\end{figure*}

In order to evaluate the photonic learning method presented in Sec.~\ref{sec:photonic_policy_learning}, we used a training environment composed of three parts, mapped to the detailed steps in Alg.~\ref{algorithm1}.
We used a photonic circuit simulator, classical optimizers, and a restricted gym environment suitable for RL as depicted in Fig.~\ref{fig:arch}.

In the continuous control cycle, the parametric policy circuit infers a set of actions $a_t$ during a trajectory rollout. The simulated environment gives feedback as reward $r_t$ and state observation vector $s_t$ at each step. The classical optimizer of the value function receives both rewards and policy feedback, while the policy optimizer considers the critic's feedback via its combined loss function. Both optimizers use stochastic gradient methods with Adam optimizer\cite{kingma2017adam}. 
During training, multiple agents are trained independently, initialized with different random seeds.

\subsubsection*{Photonic Simulator Backend}
The photonic quantum policy circuit is simulated on Strawberry Fields (SF) \cite{strawberryfields}, which
is an open-source photonic quantum programming framework. 
SF's basic Fock backend is integrated to two popular machine learning frameworks, TensorFlow (TF) \cite{tensorflow2015-whitepaper} and PennyLane \cite{bergholm2018pennylane,schuld2019evaluating}. 
In this work, we implement the numerical experiments using SF's Fock backend for a reduced number of circuit evaluations, i.e., to avoid double circuit evaluation for differentiation based on the parameter-shift rule in PennyLane over extensive training simulations.

It is important to note that simulating quantum photonics is especially hard for classical computers. The Fock-space of the system scales as $D^M$, where $D$ is the cutoff-dimension, i.e., $D-1$ the maximum number of photons allowed in each mode, and $M$ is the number of modes used. Using a small cutoff dimension $D<10$ compromises the reliability of the numerical results. Therefore, we used only two photonic modes, and hence were able to increase the cutoff-dimension to $D=16$.

\subsubsection*{Classical Optimization with TensorFlow Keras Modules}
The classical optimization consists of two parts. First, the value function forms the classical part of the hybrid agent. Second, because we are using a classical simulator of a quantum computer, the policy function gradient update is also calculated classically. When implemented on a real quantum processing unit, such hybrid models are able to leverage quantum circuit differentiation directly 
with high speed.
Keras \cite{chollet2015keras} was used to abstract ML models by using \textit{layers} and \textit{models} as building blocks of computation for stochastic gradient optimizers.

\subsubsection*{OpenAI Gym Environments}
OpenAI Gym \cite{1606.01540} is a general framework for developing and benchmarking reinforcement learning algorithms. 
It provides an interface to a collection of standardized training environments including an implementation of the \CartPole{} problem, the \CartPoleEnv{}, which can be used as black-box. Note that an episodic RL approach is used, i.e., the agent re-starts in a fresh, randomly initialized environment after each terminal state of the environment. 
Just like SF, the Gym framework uses TF backend, allowing us to train quantum-powered agents.

\definecolor{airforceblue}{rgb}{0.36, 0.54, 0.66}
\definecolor{cadetblue}{rgb}{0.37, 0.62, 0.63}
\definecolor{citrine}{rgb}{0.89, 0.82, 0.04}
\newcommand{\CST}[2]{\vbox{\colorbox{#1}{\vbox{\small #2\;}}}}
\newcommand{\CSTN}[2]{\vbox{\colorbox{#1}{\vbox{\small #2}}}}
\newcommand{\CSTQ}[1]{\CST{purple!10}{#1}}
\newcommand{\CSTCL}[1]{\CST{citrine!10}{#1}}
\newcommand{\CSTNQ}[2]{\CSTN{purple!10}{#1}}
\newcommand{\CSTNCL}[2]{\CSTN{citrine!10}{#1}}
\newcommand{\CSTOAI}[1]{\CST{cadetblue!17}{#1}}
\newcommand{\CSTCLF}[1]{\vbox{\colorbox{citrine!10}{\vbox{#1}}}}

\newcommand{\cfbox}[2]{%
    \colorlet{citrine}{.}%
    {\color{#1}%
    \fbox{\color{citrine}#2}}%
}


\colorbox{gray!5}{
\begin{minipage}[t]{0.9\linewidth}
\begingroup
\removelatexerrorr
\begin{algorithm}[H]
 \caption{\textbf{Photonic PPO($T$)}
 \\\small{The steps of RL agent training divided between three components of the hybrid architecture. The framed classical optimization step for update of policy function $\pi$ parameters can be performed via quantum circuit differentiation when run on QPUs in the future.}}
 \SetKwProg{myproc}{Procedure PPO}{}
 {\myproc{\proc{}}}
 \CSTQ{Init policy circuit $\pi$ parameters $\theta_0$}
 \CSTCL{Init value function $V^{\pi}_a$}
\ForEach(\emph{ \# hybrid training loop}){episode}{
    \CSTOAI{Init environment $E$}
    \For(\emph{ \# collect a set of trajectories}){$t = 1$ to $T$}{
        \CSTQ{Encode $s_t$ into $\pi(\theta)$ policy circuit}
        \CSTQ{Infer $a_t$ from $\braket{X_\varphi}_{t,\theta}$ measurements}
        \CSTOAI{Evaluate $a_t$ action in $E$}
        \CSTOAI{Collect states $s_t$, rewards $r_t$}
    }
    \CSTCL{Compute total discounted reward $V_{a,\textrm{targ}}^{\pi}$}
    \CSTCL{Estimate advantages $\hat A_t$ based on current $V^{\pi}_a$}

    \CSTCLF{\cfbox{purple}{Update $\pi$ by optimizing $L$:
    $\theta \leftarrow \theta_{\textrm{new}}$\;}}
    \CSTCL{Update $V^{\pi}_a$ by optimizing $L^{VF}$:~$V^{\pi}_a\leftarrow V^\pi_{a_\textrm{new}}$}



}

\label{algorithm1}
\end{algorithm}
\endgroup
\textbf{\small{Color key:}}\\
\colorbox{purple!10}{\parbox[t]{7.5cm}{\qquad\quad \small{Quantum simulator}}}
\colorbox{citrine!10}{\parbox[t]{7.5cm}{\qquad\quad \small{Classical optimizer}}}
\colorbox{cadetblue!17}{\parbox[t]{7.5cm}{\qquad\quad \small{Environment simulator}}}\\
\rule{1\textwidth}{0.8pt}
\end{minipage}
}

\begin{figure*}[h!]
\centering
\newcommand{\subfigsize}{0.43\textwidth}
\subfloat[Classical PPO - averaged agents]{\includegraphics[width=\subfigsize]{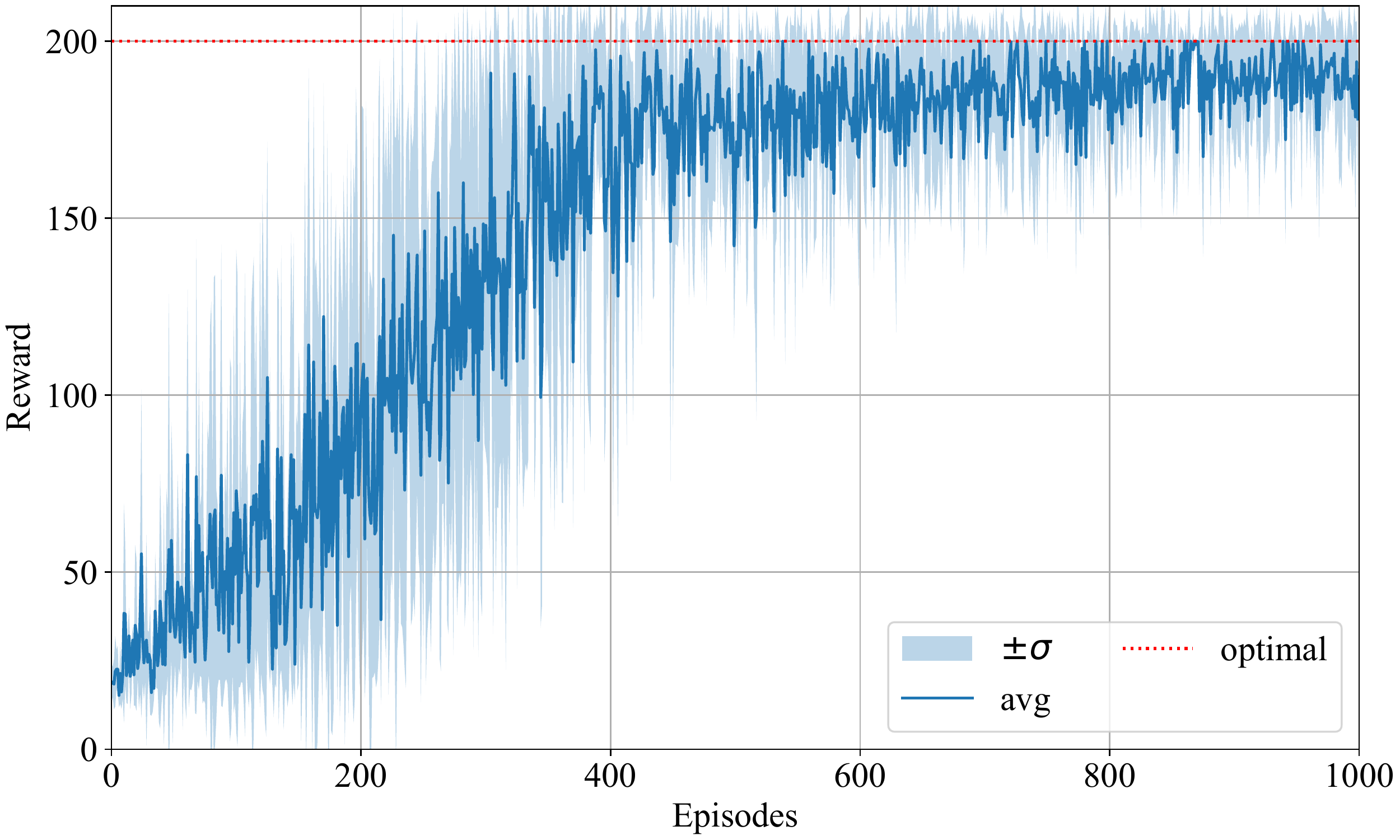}%
\label{subfig:classical_1}}
\subfloat[Classical PPO - individual agents]{\includegraphics[width=\subfigsize]{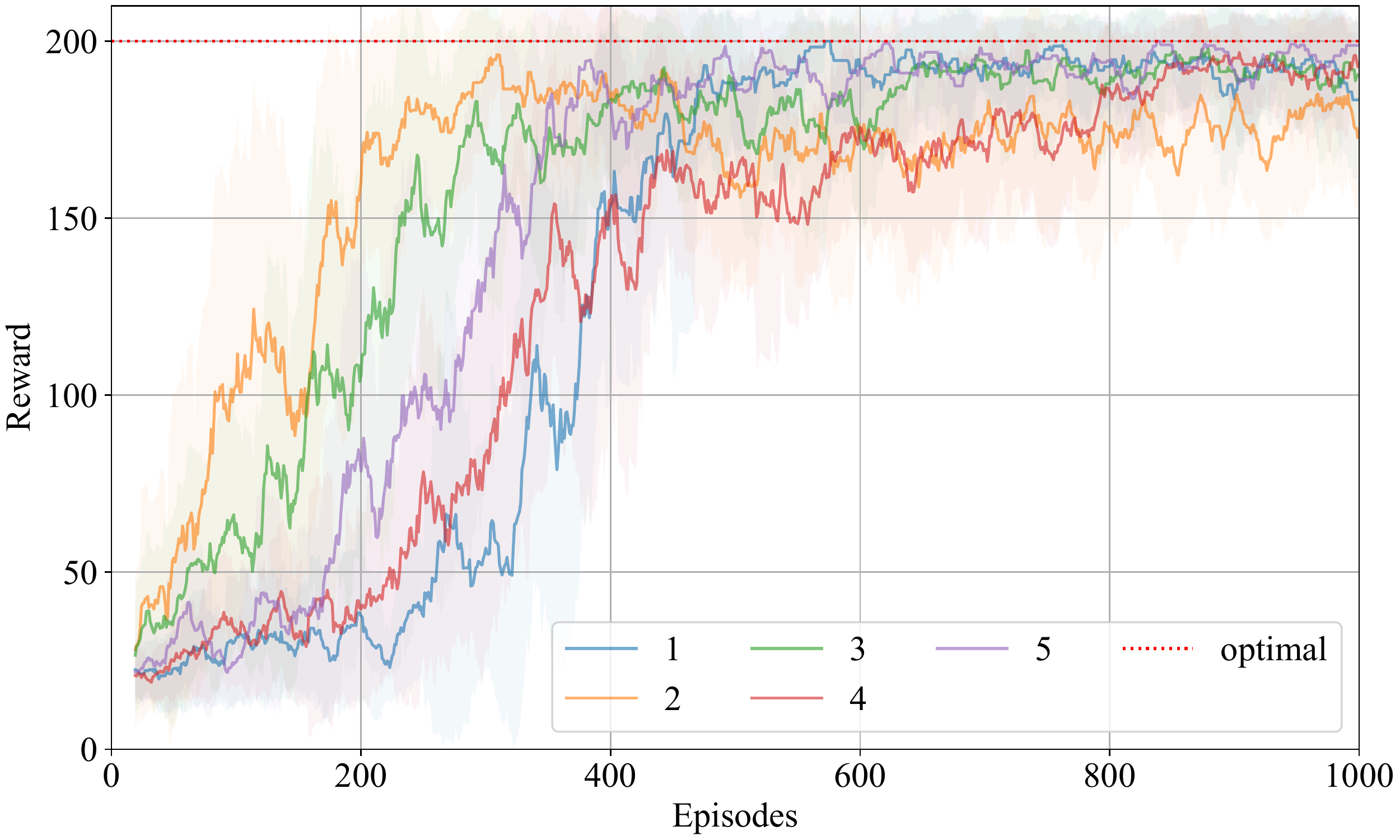}%
\label{subfig:classical_2}}

\subfloat[Photonic PPO single encoding layer - averaged agents]{\includegraphics[width=\subfigsize]{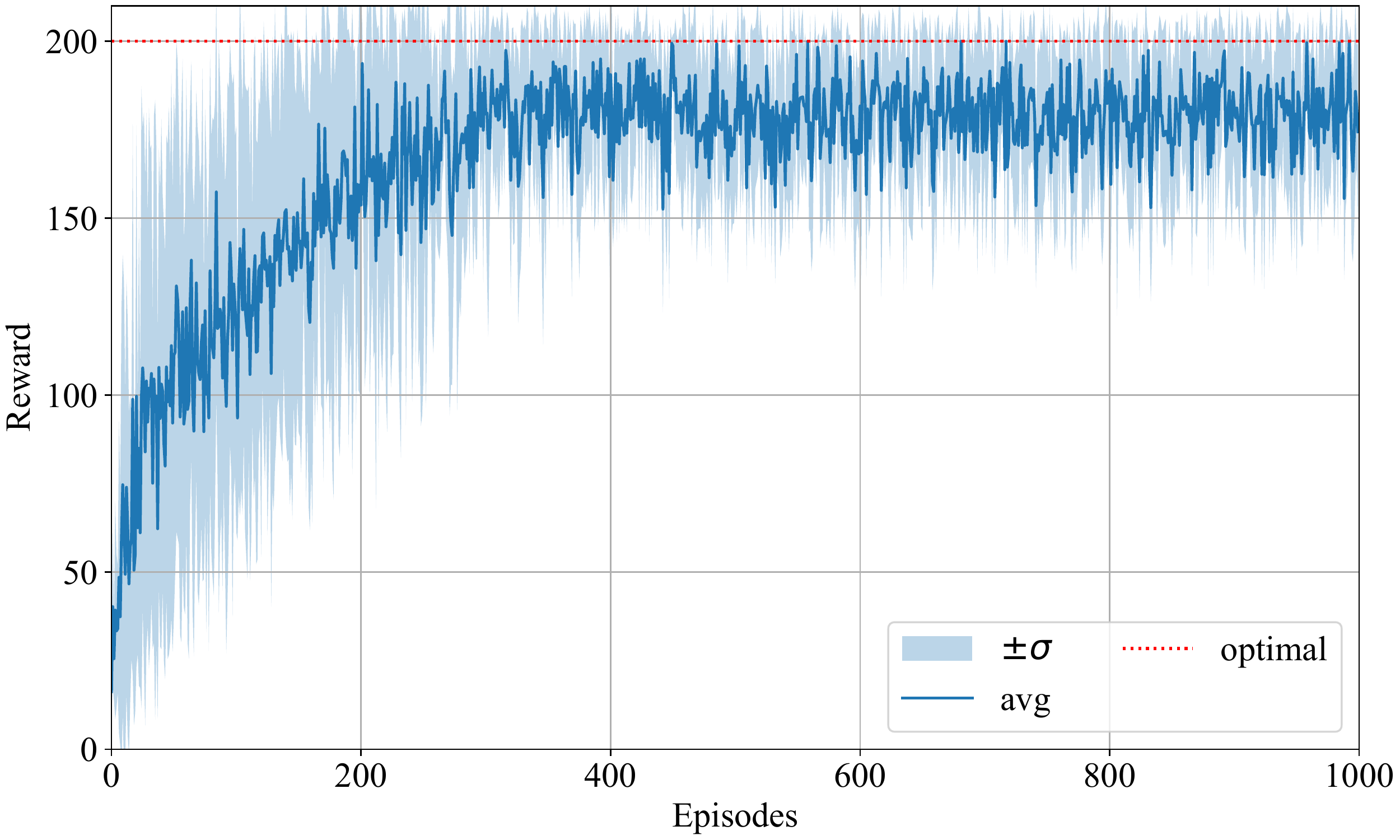}%
\label{subfig:pi32_1}}
\subfloat[Photonic PPO single encoding layer - individual agents]{\includegraphics[width=\subfigsize]{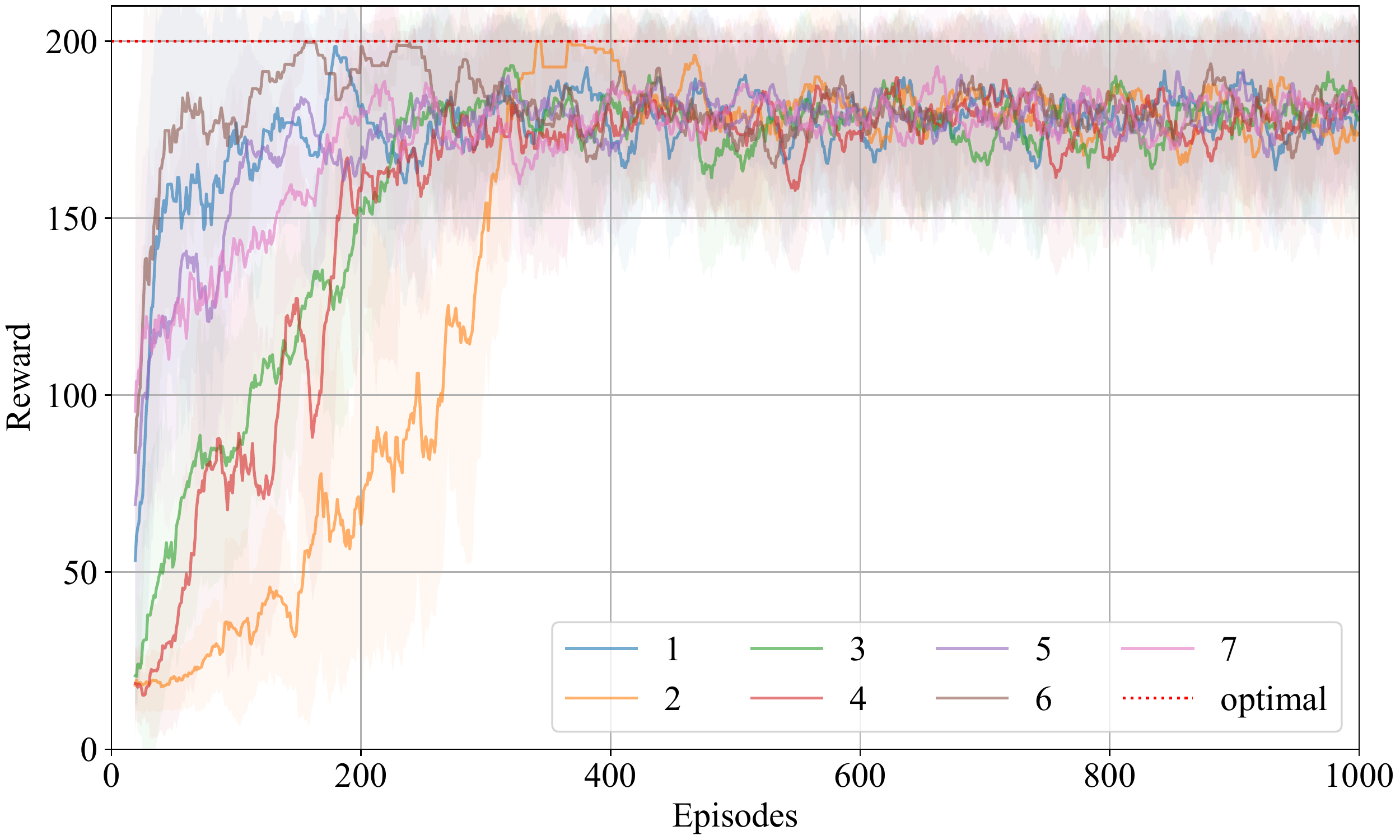}%
\label{subfig:pi32_2}}

\subfloat[Photonic PPO with data re-uploading - averaged agents]{\includegraphics[width=\subfigsize]{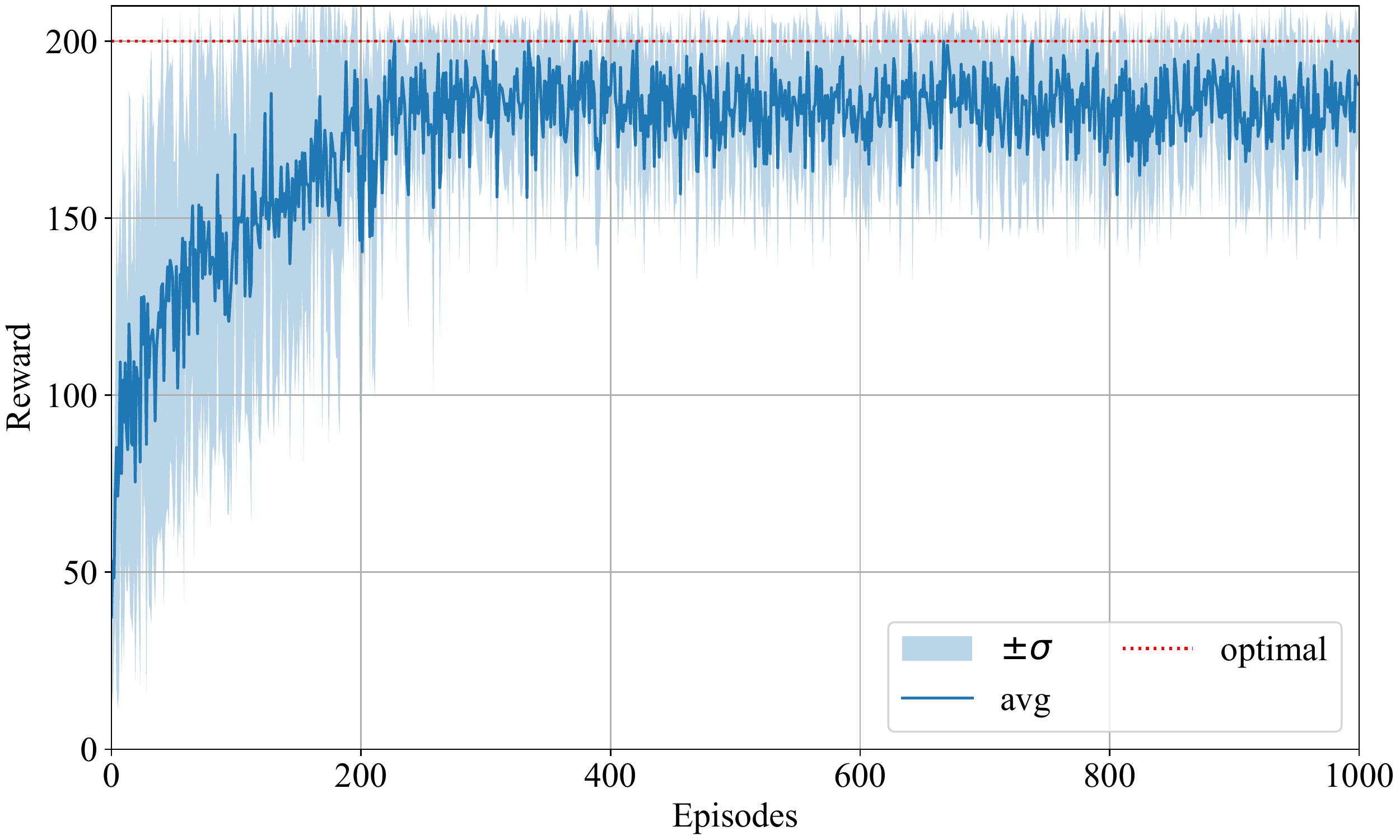}
\label{subfig:pi36_1}}
\subfloat[Photonic PPO with data re-uploading - individual agents]{\includegraphics[width=\subfigsize]{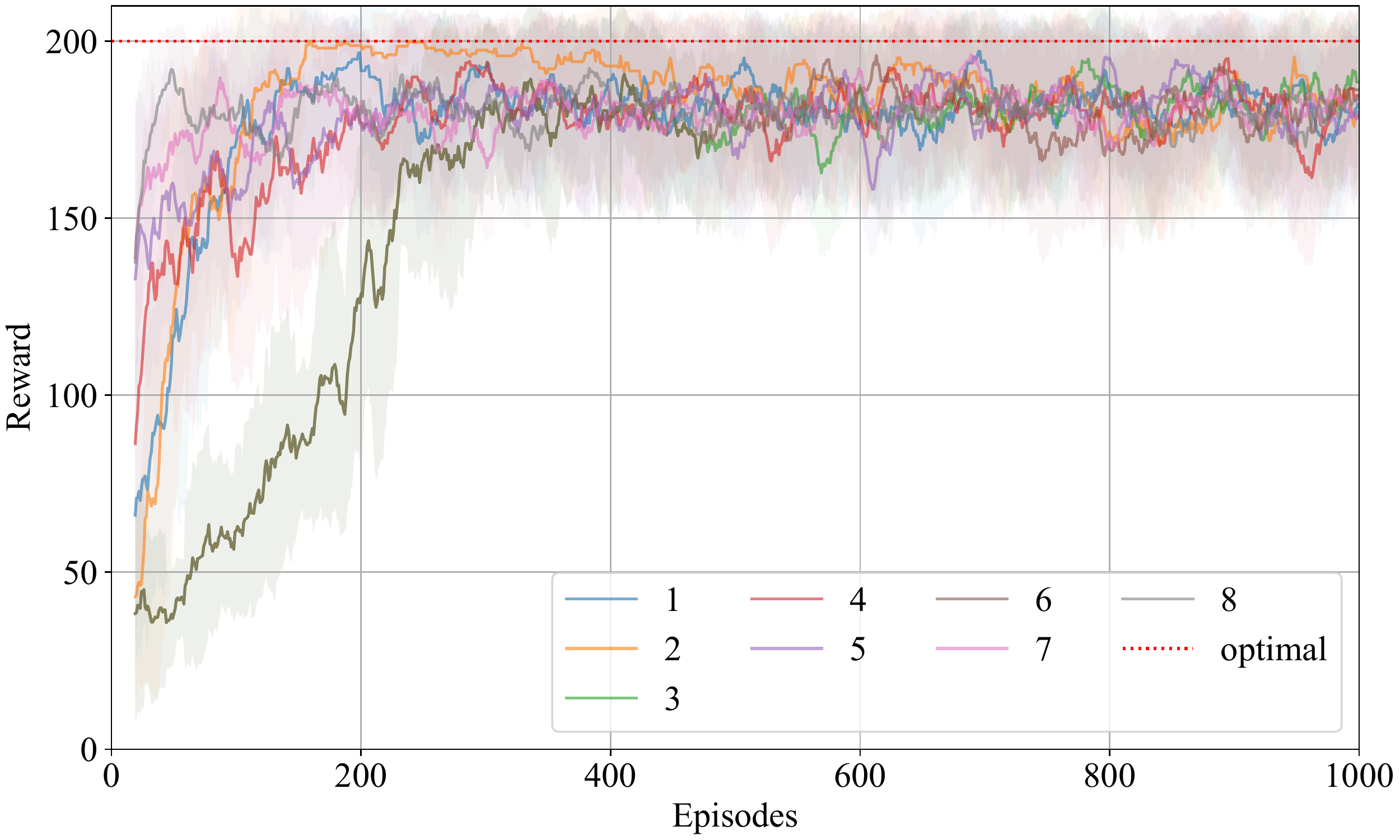}
\label{subfig:pi36_2}}
\caption{Empirical results of classical and photonic training. Training processes are shown for three types of policies. In the upper panels, Classical PPO serves as a baseline. In the middle panels, Photonic PPO with a single encoding layer shows comparable performance, while already improving upon convergence speed. In the bottom panels, Photonic PPO with data re-uploading strategy shows additional improvement on convergence speed and learning stability.
On the left, the overall averaged training curves are depicted. On the right panels, the individual agent trajectories are shown separately using a moving average of window size $20$. The corresponding standard deviations are shown with $\pm\sigma$ levels represented.
}
\label{fig:results}
\end{figure*}

\section{Empirical Results}
\label{sec:empirical_results}


\newcommand{\basic}{single-encoded\xspace}
\newcommand{\reup}{data re-upload\xspace}

To demonstrate the power of the  introduced photonic PPO
method, we designed a set of numerical experiments in the following way. First, as a baseline, we trained a classical agent in the restricted \CartPole scenario presented in Sec.~\ref{subsec:pppocart}.
Next, we tested the performance of the 
3-layer QNN  with a single data encoding layer as depicted in Fig.~\ref{subfig:qnn321a}; and finally, the \reup QNN, shown on Fig.~\ref{subfig:qnn321b}.

\subsection{Technical Description of the Numerical Experiments}

All agents, classical and quantum, were run with a fixed configuration and empirically chosen hyperparameters. 
We set $\beta$ of Eq.~\eqref{eq:combined_loss} to $0.1$,
the agent's memory size to $10\,000$ units and the initial learning rate of the agent to $0.01$. 
Due to the use of the Adam optimization method, the learning rates were subject to change during training. 
The value function $V^\pi$ was modelled with a dense layer of $32$ weights and had the initial 
learning rate of $0.005$. 
The maximal trajectory that the agents could reach, i.e., the horizon, was set to $200$. 
The clippings were done with $\epsilon = 0.2$, the entropy coefficient $c_2$ was fixed to $0.01$. 
We worked with minibatches of size $8$ and discount factor $\gamma = 0.99$ furthermore, we set the GAE parameter to $\lambda = 0.95$ and used softmax activation. The $L_2$-regularization factor was set to $\alpha = 0.075$.

For the sake of completeness, we also discuss the execution time of the simulations.
For the QNNs, the run times are influenced by the SF's  Fock backend implementation, whereas the classical NN's run time depends only on TF's performance. We used TF's CPU backend and measured $26$ minutes of CPU time on average for the classical agents $10.6$ and $16.7$ hours of CPU time for the \basic and the \reup agents for reaching $1000$ episodes, respectively. 

\subsection{Evaluation of the Numerical Results}

We trained the classical and quantum agents with a restricted version of the \CartPole{} problem, in which the agents could only observe two out of the four features of the environment, namely the pole's angle and pole's angular velocity. 
The \basic and \reup QNNs share the same set of hyperparameters and only differ in the 
architecture of their quantum circuits. 
However, both architectures result in the same number of trainable parameters of $42$. 
The structure of the classical network was chosen to match the quantum circuit in the number of trainable parameters. For this purpose, we used the simplest NN with a 2-dimensional input layer, a single dense hidden layer of size $8$, and a dense output layer of size $2$, resulting in $42$  trainable parameters.

In our experiments, we started to train $20$ agents in each setup. Two filtering rules were applied: First, we terminated the agents reaching a reward less than $100$ at episode $500$, since in RL one usually trains large number of agents, and the reward of the underperforming agents are not relevant. Second, we also terminated the agents starting with an initially good policy (starting reward higher than $100$) in order to have a better comparison of the learning procedures.

Fig.~\ref{fig:results} shows the mean and standard deviation of the reward of the surviving agents, and for demonstration purposes, the moving average and corresponding standard deviation of the reward for each individual agent. We observe a convergence with reduced variance and stable performance for the classical, the \basic and the \reup reached at Episode no.~$500$, $300$, and $200$, respectively.

Our results clearly illustrate that the \basic QNN agents learn the problem at a comparable level as their classical counterpart. In addition, the data re-uploading strategy not only kept the performance level, but also was able to improve the speed towards convergence while enhancing the stability of the training via decreased variance.

To summarize, the presented method is successful in learning the continuous control problem, and the experiments strengthen the picture emerging from a series of QML studies that data re-uploading strategy for universal quantum circuit representations improves the convergence properties of variational quantum learning.

\section{Conclusion and Outlook}
\label{sec:conclusion}

In this study, we presented a photonic quantum generalization of proximal policy optimization for RL, adapted to a simple continuous control problem. We used a training setup of three components, using standard open-source frameworks for optimizers, circuit, and environment simulators.
Our empirical results showed that the novel photonic RL agents can not only keep up with but even outperform classical ones, i.e., the presented single encoding layer solution already show superior speed in initial ramp up stage. Then, adding the data re-upload method, this became a more prominent advantage, with the optimal policy reached in half of the time necessary for the comparable sized classical solution.
In future work, we plan to adapt different RL methods, to implement an extension to higher-dimensional problem spaces, and evaluate RL methods with noise effects included in realistic photonic simulators.

\section*{Acknowledgment}
Zsolt Tabi and Zolt{\'a}n Zimbor{\'a}s would like to thank the support of the Hungarian Quantum Technology National Excellence Program (Project No. 2017-1.2.1-NKP-2017-00001), and acknowledge also support from the Hungarian National Research, Development and Innovation Office (NKFIH) within the  Quantum Information National Laboratory of Hungary and through Grants No. FK 135220, K124176, KH129601.

\bibliographystyle{IEEEtran}
\bibliography{IEEEabrv,pppo}

\end{document}